\documentclass[12pt]{iopart}
\usepackage{graphicx}  
\usepackage{color}  
\usepackage{bm}
\usepackage{url}

\begin{document}

\title[Perturbative effects of spinning black holes]
{Perturbative effects of spinning black holes with applications to recoil velocities}

\author{Hiroyuki Nakano, Manuela Campanelli, Carlos O. Lousto, Yosef Zlochower}
\address{Center for Computational Relativity and Gravitation, 
School of Mathematical Sciences, 
Rochester Institute of Technology, Rochester, New York 14623, USA}
\ead{nakano@astro.rit.edu, manuela@astro.rit.edu, lousto@astro.rit.edu, yosef@astro.rit.edu}
\begin{abstract}
Recently, we proposed an enhancement of the
Regge-Wheeler-Zerilli formalism for first-order perturbations 
about a Schwarzschild background that includes 
first-order corrections due to the background black-hole spin.
Using this formalism, we investigate gravitational wave recoil effects
from a spinning black-hole binary system analytically. 
This allows us to better understand the origin of the large recoils
observed in full numerical simulation of spinning black hole binaries.
\end{abstract}

%Uncomment for PACS numbers title message
\pacs{04.25.Nx, 04.70.Bw, 04.30.Db}
% Keywords required only for MST, PB, PMB, PM, JOA, JOB? 
%\vspace{2pc}
%\noindent{\it Keywords}: Article preparation, IOP journals
% Uncomment for Submitted to journal title message
\submitto{\CQG}
% Comment out if separate title page not required
\maketitle

%%%%%%%%%%%%%%%%%%%%%%%%%%%%%%%%%%%%%%%%%%%%%%%%%%%%%%%%%%%%%%%%%%%%%%%%
\section{Introduction}
%%%%%%%%%%%%%%%%%%%%%%%%%%%%%%%%%%%%%%%%%%%%%%%%%%%%%%%%%%%%%%%%%%%%%%%%

After the breakthroughs of 2005~\cite{Pretorius:2005gq, 
Campanelli:2005dd, Baker:2005vv} with the fully non-linear 
dynamical numerical simulation of the inspiral, merger and
ringdown of black-hole binaries (BHBs), 
there were many important advances 
in the understanding of black-hole physics.
Indeed, the discovery and modeling of very large recoil 
velocities~\cite{Campanelli:2007ew,Campanelli:2007cga} acquired by the final 
remnant of spinning BHB mergers have attracted a lot of 
interest among astrophysicists.

Empirical formulae for the final remnant black hole recoil 
velocity (also mass and spin) from merging black-hole binaries 
 were obtained in~\cite{Lousto:2009mf} (and references therein),
where post-Newtonian results~\cite{Kidder:1995zr, Racine:2008kj}
 were used as guide to model the recoil dependence on the physical
 parameters 
of the progenitor BHB~\cite{Lousto:2009mf}.
On the other hand, there is also a long history 
of recoil studies in black hole perturbation theory
(e.g., see \cite{Oohara:1983gq,Fitchett:1984qn}), and 
recently, an analytic treatment of the linear momentum 
flux of the plunge of a particle 
into a Kerr black hole has been considered in~\cite{Mino:2008at}.  

In~\cite{Lousto:2010qx}, we extended the Regge-Wheeler-Zerilli (RWZ)
equations~\cite{Regge:1957td,Zerilli:1971wd} 
for the Schwarzschild perturbations by including, perturbatively, 
a term linear in the spin of the larger black hole (SRWZ formulation).
We have found good agreement in the full numerical 
and perturbative waveforms for intermediate mass ratio black-hole binaries, 
reaching 99.5\% matching for the leading $(\ell,m)=(2,2)$ 
mode~\cite{Lousto:2010tb,Lousto:2010qx} 
and recently simulated a mass ratio
$100:1$ BHB merger~\cite{Lousto:2010ut}. 
This formalism can be considered as a extension 
of \cite{Gleiser:2001in} in the close limit.
Here, using the slow motion approximation, 
we derive the evolution of the linear momentum for binary systems 
in the SRWZ formalism analytically and compare to the post-Newtonian
(PN) expansions.

%%%%%%%%%%%%%%%%%%%%%%%%%%%%%%%%%%%%%%%%%%%%%%%%%%%%%%%%%%%%%%%%%%%%%%%%
\section{Formulation}
%%%%%%%%%%%%%%%%%%%%%%%%%%%%%%%%%%%%%%%%%%%%%%%%%%%%%%%%%%%%%%%%%%%%%%%%

%%%%%%%%%%%%%%%%%%%%%%%%%%%%%%
\subsection{Spin as a perturbation}\label{sec:SAAP}
%%%%%%%%%%%%%%%%%%%%%%%%%%%%%%

Our goal is to analytically model the waveforms from
 a particle with mass $\mu$ orbiting around a
spinning black hole with mass $M$ (where $\mu \ll M$). 
In~\cite{Lousto:2010qx}, we considered 
the Kerr metric up to $O(a^1)$, where $a$ denotes the spin 
of the BH which has the dimension of mass, 
and the spin direction is along the $z$-axis. 
In this paper, we set the spin along the $x$-axis, 
and the metric is given by 
\begin{eqnarray}
\fl
ds^2 = - \left(1-\frac{2M}{r}\right)dt^2
+\left(1-\frac{2M}{r}\right)^{-1} dr^2 
+ r^2 \left( d\theta^2+\sin^2\theta d\phi^2\right) 
\nonumber \\  
+\frac{4Ma}{r}\,
dt\left(\sin\phi d\theta + \sin\theta\cos\theta\cos\phi d\phi \right)
+ O(a^2) \,. 
\label{eq:Sch+spin}
\end{eqnarray}
in the Boyer-Lindquist radial coordinate. 
The last term in the right hand side of the above equation  
is treated as a perturbation in the Schwarzschild black hole background. 
\begin{eqnarray}
\fl
g_{\mu\nu} = g_{\mu\nu}^{\rm Sch} + h_{\mu\nu}^{\rm (1,spin)}
\,; \nonumber \\
\fl
h_{t\theta}^{\rm (1,spin)} = h_{\theta t}^{\rm (1,spin)} = \frac{2S_x}{r}\sin\phi \,,
\quad 
h_{t\phi}^{\rm (1,spin)} = h_{\phi t}^{\rm (1,spin)} 
= \frac{2S_x}{r}\sin\theta\cos\theta\cos\phi \,,
\label{eq:htth-htph}
\end{eqnarray}
where $S_x=M a$. 
By using the expansion defined by 
the tensor harmonics given in~\cite{Nakano:2007cj}, 
we find that the coefficients of the tensor harmonics are given by
\begin{eqnarray}
h_{0\,11}^{\rm (1,spin)}(t,r) = -\sqrt{\frac{8\pi}{3}}\,\frac{S_x}{r} \,,
\quad 
h_{0\,1-1}^{\rm (1,spin)}(t,r) &=& \sqrt{\frac{8\pi}{3}}\,\frac{S_x}{r} \,.
\label{eq:h0S}
\end{eqnarray}

%%%%%%%%%%%%%%%%%%%%%%%%%%%%%%
\subsection{SRWZ formulation}\label{sec:SRWZ}
%%%%%%%%%%%%%%%%%%%%%%%%%%%%%%

We treat the coupling between the spin discussed above 
and first order metric perturbation (gravitational radiation) 
as a second order perturbation calculation. 
The Einstein equations up to the second perturbative order are formally 
written as  
\begin{eqnarray}
G_{\mu\nu}^{(1)}[h^{(1)}] 
+ G_{\mu\nu}^{(1)}[h^{(2)}]
+ G_{\mu\nu}^{(2)}[h^{(1)},h^{(1)}] 
= 8\,\pi\,T_{\mu\nu}
\,,
\end{eqnarray}
where the energy-momentum tensor $T_{\mu\nu}=T_{\mu\nu}^{(1)} + T_{\mu\nu}^{(2)}$, 
and $h_{\mu\nu}^{(1)}$ and $h_{\mu\nu}^{(2)}$ are the 
first and second order metric perturbations, respectively. 
We consider that the second order metric perturbation, $h^{\rm (2,wave)}$
is created by the spin $h^{\rm (1,spin)}$ and radiation $h^{\rm (1,wave)}$ couplings. 
In this case, we solve 
\begin{eqnarray}
G_{\mu\nu}^{(1)}[h^{\rm (1,wave)}] &=& 8\,\pi\,T_{\mu\nu} \,,
\label{eq:formal1stE}
\\
G_{\mu\nu}^{(1)}[h^{\rm (2,wave)}] &=& - G_{\mu\nu}^{(2)}
[h^{\rm (1,wave)},h^{\rm (1,spin)}] \,,
\label{eq:formal2ndE}
\end{eqnarray}
up to $O(a^1)$. 
Here the square of the first order radiation has been ignored.  
We solve the above equations in the tensor harmonics expansion 
as in the RWZ formalism.

%%%%%%%%%%%%%%%%%%%%%%%%%%%%%%
\subsection{Gravitational radiation and linear momentum evolution}
%%%%%%%%%%%%%%%%%%%%%%%%%%%%%%

Gravitational wave modes in the RWZ formalism 
are given by the metric perturbation 
under an asymptotic flat (AF) gauge
\begin{eqnarray}
{\bf h}^{(n)} =& \sum_{\ell m} 
\left[
\left[{\frac{1}{2}\ell(\ell+1)(\ell-1)(\ell+2)}\right]^{1/2}
G^{(n) {\rm AF}}_{\ell m}(t,r)\,{\bf f}_{\ell m} 
\right. \nonumber \\ & \left. 
+ \frac{\left[{2\ell(\ell+1)(\ell-1)(\ell+2)}\right]^{1/2}}{2r^2}
\,h^{(n) {\rm AF}}_{2\,\ell m}(t,r)\,{\bf d}_{\ell m} 
\right]\,, 
\label{eq:MPinAF}
\end{eqnarray}
where superscripts $(n)$ ($n=1,\,2$) denote the perturbative order, 
and ${\bf f}_{\ell m}$ and ${\bf d}_{\ell m}$ are tensor harmonics. 
The even parity mode $G^{(n) {\rm AF}}_{\ell m}$ and 
odd parity mode $h^{(n) {\rm AF}}_{2\,\ell m}$ 
are expressed by the Regge-Wheeler-Zerilli functions 
$\psi_{\ell m}^{(n){\rm (even)}}$ and $\psi_{\ell m}^{(n){\rm (odd)}}$ as follows. 
\begin{eqnarray}
G^{(n) {\rm AF}}_{\ell m}(t,r) = \frac{1}{r}\, \psi_{\ell m}^{(n){\rm (even)}}(t,r) \,,
\quad 
h^{(n) {\rm AF}}_{2\,\ell m}(t,r) = i\,r \,\psi_{\ell m}^{(n){\rm (odd)}}(t,r) \,.
\end{eqnarray} 
(See the definition of the tensor harmonics expansion 
given in \cite{Nakano:2007cj}, and \cite{Lousto:2010qx} 
for more precise discussion of the second perturbative order.)

Next, we discuss the time evolutions of the linear momentum of the binary system. 
This is obtained from the following expression 
(e.g. see Eq.~(2.17) in \cite{Fitchett:1984qn}).  
\begin{eqnarray}
{\dot P}_i &=& -  \frac{1}{32\pi} 
\int d\Omega \,r^2 
\,n_i \,
\, \left< h^{\alpha\beta}{}_{;t}h_{\alpha\beta;t} \right>_{\rm TT}
\,, 
\label{eq:Pxyz}
\end{eqnarray}
where $n_i=(\sin\theta\cos\phi,\,\sin\theta\sin\phi,\,\cos\theta)$.   
The subscript TT means the transverse traceless part and 
we may use the metric in Eq.~(\ref{eq:MPinAF}) under the AF gauge. 
Then, the above equation is written as 
\begin{eqnarray}
\fl
{\dot P}_i = -  \frac{1}{64\pi} 
\sum_{\ell m} \sum_{\ell'm'} 
\nonumber \\ 
\fl \qquad \qquad 
\biggl[ 
\left(
r^2\,\dot G^{(n) {\rm AF}}_{\ell m}(t,r) 
\,\dot G^{(n') {\rm AF}}_{\ell' m'}(t,r)
-\frac{1}{r^2}\, \dot h^{(n) {\rm AF}}_{2\,\ell m}(t,r) 
\,\dot h^{(n') {\rm AF}}_{2\,\ell' m'}(t,r) \right) P^{\rm S}_i(\ell m \ell' m')
\nonumber \\
\fl \qquad \qquad \quad 
+ i\,\left(
\dot G^{(n) {\rm AF}}_{\ell m}(t,r) 
\,\dot h^{(n') {\rm AF}}_{2\,\ell' m'}(t,r)
- \dot h^{(n) {\rm AF}}_{2\,\ell m}(t,r) 
\,\dot G^{(n') {\rm AF}}_{\ell' m'}(t,r) \right) P^{\rm C}_i(\ell m \ell' m') \biggr] \,,
\label{eq:dotPi}
\end{eqnarray}
where 
\begin{eqnarray}
P^{\rm S}_i(\ell m \ell' m')&=& 
\int d\Omega 
\,n_i \,
\left(
W_{\ell m} W_{\ell' m'} + \frac{1}{\sin^2\theta}X_{\ell m} X_{\ell' m'} 
\right) \,, 
\nonumber \\ 
P^{\rm C}_i(\ell m \ell' m')&=& 
\int d\Omega 
\,n_i \, 
\frac{1}{\sin \theta}\left(
W_{\ell m} X_{\ell' m'} - X_{\ell m} W_{\ell' m'}
\right) \,.
\label{eq:PsPc}
\end{eqnarray}
We note that $P^{\rm S}_i$ vanishes for $\ell-\ell'=2k$ ($k$: integer) 
because of the parity of the integration, 
while $P^{\rm C}_i$ vanishes for $\ell-\ell'=2k+1$ ($k$: integer). 

In the following, we discuss only the leading order contribution 
of gravitational waves to the linear momentum evolution 
in the slow motion approximation. 
In practice, the combination of the first order $\ell=2$, $m=0$ 
and $\ell=3$, $m=0$ even parity modes produces 
the leading order contribution for the non-spinning case, 
and the combination of $\ell=2$, $m=0$ even parity 
and the $\ell=2$, $m=\pm1$ odd parity modes becomes 
the dominant contribution of spin in the situation discussed below.

%%%%%%%%%%%%%%%%%%%%%%%%%%%%%%%%%%%%%%%%%%%%%%%%%%%%%%%%%%%%%%%%%%%%%%%%
\section{Evolution of the linear momentum}
%%%%%%%%%%%%%%%%%%%%%%%%%%%%%%%%%%%%%%%%%%%%%%%%%%%%%%%%%%%%%%%%%%%%%%%%

%%%%%%%%%%%%%%%%%%%%%%%%%%%%%%
\subsection{Point particle's motion}\label{subsec:PPM}
%%%%%%%%%%%%%%%%%%%%%%%%%%%%%%

We consider a particle falling radially 
into a Schwarzschild black hole as the first order source. 
Assuming $\Theta(\tau)=\Phi(\tau)=0$
\footnote{Although we write $\Phi(\tau)=0$ here, 
$\Phi(\tau)$ is arbitrary at $\Theta(\tau)=0$.},
where the particle's location 
is given by $\{T(\tau),\,R(\tau),\,\Theta(\tau),\,\Phi(\tau) \}$, 
the equation of motion of the particle is
\begin{eqnarray}
\left(\frac{dR}{dt}\right)^2 &=& -\left(1-\frac{2M}{R}\right)^3 \frac{1}{E^2}
+ \left(1-\frac{2M}{R}\right)^2 \,,
\end{eqnarray}
where $R=R(t)$ is the location of the particle 
and the energy $E$ is written by 
\begin{eqnarray}
E &=& 
\left(1-\frac{2M}{R} \right)\,\frac{dT(\tau)}{d\tau} \,.
\end{eqnarray}
We also use 
\begin{eqnarray}
\frac{d^2R}{dt^2} &=& 
-\frac{3}{E^2}\left(1-\frac{2M}{R}\right)^2 \frac{M}{R^2}
+ 2\left(1-\frac{2M}{R}\right) \frac{M}{R^2} \,, 
\end{eqnarray}
to simplify the calculations below. In the slow motion approximation 
where we consider $dR/dt \ll 1$ and $M/R \ll 1$, we have 
$E=1$ and $d^2R/dt^2 = -M/R^2$.

The tensor harmonics coefficients of the first order 
stress-energy tensor become
\begin{eqnarray}
{\cal A}^{(1)}_{\ell m}(t,r)
&=&
\mu \,\displaystyle{\frac{E\,R(t)}{R(t)-{2M}}}\, 
\left({dR \over dt}\right)^2 \frac{1}{(r-2M)^2}
\,\delta(r-R(t))\,Y_{\ell m}^*\left(0,0\right) \,, 
\nonumber \\ 
{\cal A}^{(1)}_{0\,\ell m}(t,r)
&=& \mu \,\displaystyle{\frac{E\,R(t)}{R(t)-{2M}}}\, \frac{(r-2M)^2}{r^4} 
\,\delta(r-R(t))\,Y_{\ell m}^*\left(0,0\right) \,,
\nonumber \\ 
{\cal A}^{(1)}_{1\,\ell m}(t,r)
&=& \sqrt{2}\,i\,\mu \,\displaystyle{\frac{E\,R(t)}{R(t)-{2M}}}\, {dR \over dt} \frac{1}{r^2}
\,\delta(r-R(t))\,Y_{\ell m}^*\left(0,0\right) \,.
\label{eq:source}
\end{eqnarray}
Otherwise the coefficients are zero. 
Here, only the $m=0$ modes have a non-zero value. 

We note that the black-hole spin effect in the equations of motion 
of the particle does not contribute to leading order 
in the slow motion approximation. Therefore, we do not consider 
any effect of the background spin on the particle's trajectory in this paper.

%%%%%%%%%%%%%%%%%%%%%%%%%%%%%%
\subsection{First order $\ell=2,\,m=0$ 
and $\ell=3,\,m=0$, even parity perturbation}\label{subsec:FO2030E}
%%%%%%%%%%%%%%%%%%%%%%%%%%%%%%

In order to calculate the radiative even parity modes, 
we use the Zerilli equation. 
For example, the Zerilli equation for the $\ell=2,\,m=0$ mode 
is given by  
\begin{eqnarray}
\fl
\left[
-{\frac {\partial ^{2}}{\partial {t}^{2}}}
+\frac{\partial^2}{{\partial r^*}^2} 
-6\,{\frac {( r-2\,M) 
( 4\,{r}^{3} +4\,{r}^{2}M+6\,r{M}^{2}+3\,{M}^{3}) }{{r}^{4} ( 2\,r+3\,M ) ^{2}}}
\right]\,\psi^{(1){\rm (even)}}_{20}(t,r) 
%\nonumber \\ 
%\fl 
%=-8\,\pi\,{\frac { \mu \, 
%\left( 2\,{R(t)}^{2}-2\,R(t){E}^{2}M+6\,R(t)M+{M}^{2} \right) 
%\left( R(t)-2\,M \right)^2  }{E {R(t)}^{3}\left( 2\,{R(t)}+3\,M \right) ^{2}}}
%Y_{2 0}^*\left(0,0\right) 
%\delta \left( r-R(t)  \right)
%\nonumber \\ 
%\fl
%+\frac{8\,\pi }{3}\,{\frac { \mu\left( R(t)-2\,M \right) ^{3} }
%{E{R(t)}^{2} \left( 2\,R(t)+3\,M \right) }}
%Y_{2 0}^*\left(0,0\right)\frac{d }{dr}\delta \left( r-R(t)  \right) 
\nonumber \\ 
= \left[ -{\frac {4\,\pi\, \mu}{{R(t)}}} 
\delta \left( r-R(t)  \right)
+\frac{4\,\pi \,\mu}{3}
\frac{d }{dr}\delta \left( r-R(t)  \right) 
\right] \,Y_{2 0}^*\left(0,0\right) 
\,.
\end{eqnarray}
In the right hand side of the above equation, 
we used the slow motion approximation, 
i.e., the characteristic orbital velocity $v \ll 1$ 
where the particle's velocity $dR(t)/dt \sim v$ 
and the potential $M/R(t) \sim v^2$, and only consider
leading-order terms in this approximation. 
%\begin{eqnarray}
%Y_{20}(\theta,\phi) = \sqrt{\frac{5}{4\pi}} 
%\left(\frac{3}{2}\cos^2 \theta -\frac{1}{2} \right)\,.
%\end{eqnarray}

The Zerilli function is obtained by the Green's function method. 
In practice, we used the Fourier transformation and prepared 
the Green's function in the frequency domain. We consider only 
the outside Green's function of the particle's location 
in the $M \to 0$ limit. This is given by 
\begin{eqnarray}
G(r,\,r') = i \, \omega \, j_2(\omega r') \,h_2^{(1)}(\omega r) \, 
\theta (r-r') \,.
\end{eqnarray}
Here, we have asymptotic behaviors, 
$j_2(\omega r)=(\omega r)^2/15$ for small $r$, and 
$h_2^{(1)}(\omega r) = i\exp(i \omega r)/(\omega r)$ for large $r$. 
Finally, we obtain 
\begin{eqnarray}
\psi^{(1){\rm (even)}}_{20}(t,r) &=&
\frac{16\, \pi}{15} \mu
\left(\dot R^2 - \frac{M}{R} \right) 
Y_{2 0}^*\left(0,0\right) \,,
%&=& \frac{16\, \pi}{15} \mu
%\left(\dot R^2 + R \ddot R \right) 
%Y_{2 0}^*\left(0,0\right) \,,
\label{eq:psi20e}
\end{eqnarray}
where the argument of $R$ is the retarded time $(t-r)$. 

In the same way, we calculate the $\ell=3,\,m=0$ even parity mode. 
In the leading order of the slow motion approximation, 
the source term of the Zerilli equation becomes 
%\begin{eqnarray}
%\left[
%-{\frac {\partial ^{2}}{\partial {t}^{2}}}
%+\frac{\partial^2}{{\partial r}^2} 
%-\frac{12}{r^2}
%\right]\,\psi^{(1)(even)}_{30}(t,r) 
%= S_{30}^{(1)(even)} \,,
%\end{eqnarray}
%where the source term of the Zerilli equation is given by 
\begin{eqnarray}
S_{30}^{(1){\rm (even)}}
= \left[ - \frac{8\,\pi\,\mu }{5\,R(t)}
\delta \left( r-R(t)  \right)
+\frac{4\,\pi \,\mu}{15}
\frac{d }{dr}\delta \left( r-R(t)  \right) 
\right] \, Y_{3 0}^*\left(0,0\right) 
\,.
\end{eqnarray}
%Here, the spherical harmonics $Y_{3 0}$ is given by
%\begin{eqnarray}
%Y_{30}(\theta,\phi) = \sqrt{\frac{7}{4\pi}} \cos \theta
%\left(\frac{5}{2}\cos^2 \theta -\frac{3}{2} \right)\,.
%\end{eqnarray}
Then, using the Green's function method with asymptotic behaviors, 
$j_3(\omega r)=(\omega r)^3/105$ for small $r$, and 
$h_3^{(1)}(\omega r) = \exp(i \omega r)/(\omega r)$ for large $r$. 
we obtain 
\begin{eqnarray}
\psi^{(1){\rm (even)}}_{30}(t,r) = \frac{16\, \pi}{105} \mu
\left(\dot R^3 - \frac{2M}{R} \dot R \right) 
Y_{3 0}^*\left(0,0\right) \,. 
\label{eq:psi30e}
\end{eqnarray}

%%%%%%%%%%%%%%%%%%%%%%%%%%%%%%
\subsection{First order dipole, even parity perturbation}\label{subsec:L1M1E}
%%%%%%%%%%%%%%%%%%%%%%%%%%%%%%

Although the first order dipole ($\ell=1$) even parity mode 
in the vacuum regions can be completely eliminated 
in the center of mass coordinate system, 
the metric perturbation is not globally pure gauge~\cite{Detweiler:2003ci}. 
The coupling between this mode and the black-hole spin creates 
the leading order spin effect on the linear momentum evolution. 
We therefore have to include the $\ell=1$ mode contributions. 

For this mode, the metric perturbations are given by 
\begin{eqnarray}
\fl
{\bf h}^{(1)}_{10} = \left(1-\frac{2M}{r}\right)H^{(1)}_{0\,10}(t,r)\,{\bf a}_{0\,10}
- \sqrt{2}\,i\,H^{(1)}_{1\,10}(t,r)\,{\bf a}_{1\,10}
+\left(1-\frac{2M}{r}\right)^{-1}H^{(1)}_{2\,10}(t,r)\,{\bf a}_{10}
\nonumber \\ 
-\frac{2\,i}{r}\,h^{(1)(e)}_{0\,10}(t,r)\,{\bf b}_{0\,10}
+\frac{2}{r}\,h^{(1)(e)}_{1\,10}(t,r)\,{\bf b}_{10}
+\sqrt{2}\,K^{(1)}_{10}(t,r)\,{\bf g}_{10} 
\,.
\end{eqnarray}
The generators of the gauge transformation are by,
\begin{eqnarray}
x^{\mu} \to x^{\mu} + \xi^{(1)\mu}_{\ell=1} \left(x^{\alpha}\right) \,;
\nonumber \\ 
\xi^{(1)\mu}_{\ell=1} =
\biggl\{ V_0^{(1)}(t,r) Y_{10}(\theta,\phi),\,V_1^{(1)}(t,r) Y_{10}(\theta,\phi),\,
\nonumber \\ \qquad \qquad
V_2^{(1)}(t,r) \partial_{\theta} Y_{10}(\theta,\phi),\,
V_2^{(1)}(t,r) \frac{\partial_{\phi} Y_{10}(\theta,\phi)}{\sin^2 \theta}
\biggl\} \,,
\label{eq:ggt1}
\end{eqnarray}
where $V_0^{(1)}$,  $V_1^{(1)}$ and  $V_2^{(1)}$ are 
three degrees of gauge freedom in the $\ell=1$ mode. 
%\begin{eqnarray}
%Y_{10}(\theta,\phi) = \sqrt{\frac{3}{4\pi}} \cos \theta \,.
%\end{eqnarray}
The metric perturbations transform under the above gauge 
transformation from a gauge (G) to a gauge (G') as 
\begin{eqnarray}
\fl
H_{0\,10}^{(1){\rm G'}}(t,r) = H_{0\,10}^{(1){\rm G}}(t,r)
+ 2\,{\frac {\partial }{\partial t}}V_0^{(1){\rm G \to G'}} \left( t,r \right) 
+2\,{\frac {M}{r \left( r-2\,M \right) }}V_1^{(1){\rm G \to G'}} \left( t,r \right)  \,, 
\nonumber \\ 
\fl
H_{1\,10}^{(1){\rm G'}}(t,r) = H_{1\,10}^{(1){\rm G}}(t,r)
+ {\frac { r-2\,M  }{r}}{\frac {\partial }{\partial r}}V_0^{(1){\rm G \to G'}}
 \left( t,r \right) -{\frac {r }{r-2\,M}}{\frac {\partial }{\partial t}}
V_1^{(1){\rm G \to G'}} \left( t,r \right) \,,
\nonumber \\ 
\fl
H_{2\,10}^{(1){\rm G'}}(t,r) = H_{2\,10}^{(1){\rm G}}(t,r)
-2\,{\frac {\partial }{\partial r}}V_1^{(1){\rm G \to G'}} \left( t,r \right) 
+2\,{\frac {M }{r \left( r-2\,M \right) }}V_1^{(1){\rm G \to G'}} \left( t,r \right) \,,
\nonumber \\ 
\fl
K_{10}^{(1){\rm G'}}(t,r) = K_{10}^{(1){\rm G}}(t,r) 
-{\frac {2}{r}}V_1^{(1){\rm G \to G'}} \left( t,r \right) 
+ 2V_2^{(1){\rm G \to G'}} \left( t,r \right)  \,,
\nonumber \\ 
\fl
h_{0\,10}^{(1){\rm (e)G'}}(t,r) = h_{0\,10}^{(1){\rm (e)G}}(t,r)
+ {\frac { r-2\,M  }{r}}V_0^{(1){\rm G \to G'}} \left( t,r \right)
-{r}^{2}{\frac {\partial }{\partial t}}V_2^{(1){\rm G \to G'}} \left( t,r \right) \,,
\nonumber \\ 
\fl
h_{1\,10}^{(1){\rm (e)G'}}(t,r) = h_{1\,10}^{(1){\rm (e)G}}(t,r)
-{\frac {r }{r-2\,M}}V_1^{(1){\rm G \to G'}} \left( t,r \right)
-{r}^{2}{\frac {\partial }{\partial r}}V_2^{(1){\rm G \to G'}} \left( t,r \right) 
 \,.
\end{eqnarray}

When we choose the gauge so that 
$h_{0\,10}^{(1){\rm (e)Z}}=h_{1\,10}^{(1){\rm (e)Z}}=K_{10}^{(1){\rm Z}}=0$, 
where the suffix Z stands for the Zerilli gauge~\cite{Zerilli:1971wd}, 
we obtain the metric perturbations
\begin{eqnarray}
\fl
H_{0\,10}^{(1){\rm Z}}(t,r) 
=  \frac{8\pi\mu E}{3M(r-2M)^2}
\left(r^3\frac{d^2R(t)}{dt^2} +M(R(t)-2M) \right) \,
\theta(r-R(t)) \,Y_{10}^{*}(0,0) \,,
\nonumber \\ 
\fl
H_{1\,10}^{(1){\rm Z}}(t,r) 
=  - \frac{8\pi\mu E\,r}{(r-2M)^2}\frac{dR(t)}{dt} \,
\theta(r-R(t)) \,Y_{10}^{*}(0,0) \,,
\nonumber \\ 
\fl
H_{2\,10}^{(1){\rm Z}}(t,r) 
=  \frac{8\pi\mu E}{(r-2M)^2}(R(t)-2M) \,
\theta(r-R(t)) \,Y_{10}^{*}(0,0) \,.
\end{eqnarray}
We note that the metric perturbations are not under the AF gauge. 
For the above metric perturbation, if we consider the gauge transformation, 
\begin{eqnarray}
\fl
V_0^{(1){\rm Z \to D}}(t,r) =
-\frac{4\pi \mu E }{3M}\,{\frac {{r}^{3} }{(r-2M)^{2}}}
{\frac {dR(t)}{dt}} \theta(r-R(t)) \,Y_{10}^{*}(0,0) \,,
\nonumber \\ 
\fl
V_1^{(1){\rm Z \to D}}(t,r) = 
-\frac{4\pi \mu E }{3M}\,{\frac { r 
 }{(r-2M)}}\left(R(t) - 2\,M \right)
\theta(r-R(t)) \,Y_{10}^{*}(0,0) \,,
\nonumber \\ 
\fl
V_2^{(1){\rm Z \to D}}(t,r) = 
-\frac{4\pi \mu E }{3M}\,{\frac { 1
 }{\left( r-2\,M \right) }}\left(R(t) - 2\,M \right)  
\theta(r-R(t)) \,Y_{10}^{*}(0,0) \,,
\end{eqnarray}
we obtain the singular metric perturbation at the particle's location. 
\begin{eqnarray}
H_{0\,10}^{(1){\rm D}}(t,r) &=& 
\frac{8\pi\mu E}{3M}
\frac{R(t)^3}{(R(t)-2M)^2} \left(\frac{dR(t)}{dt}\right)^2 \,
\delta(r-R(t)) \,Y_{10}^{*}(0,0) \,,
\nonumber \\ 
H_{1\,10}^{(1){\rm D}}(t,r) &=& 
- \frac{8\pi\mu E}{3M}
\frac{R(t)^2}{R(t)-2M} \frac{dR(t)}{dt} \,
\delta(r-R(t)) \,Y_{10}^{*}(0,0) \,,
\nonumber \\ 
H_{2\,10}^{(1){\rm D}}(t,r) &=& 
\frac{8\pi\mu E}{3M}
R(t) \,\delta(r-R(t)) \,Y_{10}^{*}(0,0) \,,
\nonumber \\ 
h_{0\,10}^{(1){\rm (e)D}}(t,r) &=& 
-\frac{4\pi\mu E}{3M}
R(t)^2 \frac{dR(t)}{dt} \,
\delta(r-R(t)) \,Y_{10}^{*}(0,0) \,,
\nonumber \\ 
h_{1\,10}^{(1){\rm (e)D}}(t,r) &=& 
\frac{4\pi\mu E}{3M}
R(t)^2 \,
\delta(r-R(t)) \,Y_{10}^{*}(0,0) \,,
\nonumber \\ 
K_{10}^{(1){\rm D}}(t,r) &=& 0 \,,
\end{eqnarray}
The above means that although the metric perturbations 
vanish in the vacuum regions, 
there are some contributions at the location of the particle. 
This gauge choice has been discussed in \cite{Detweiler:2003ci} 
for circular orbit. 
The coordinate system under this gauge condition can be considered as 
the center of mass system which is suitable for the analysis 
of the evolution of the linear momentum. 
Therefore, we use this gauge and its metric perturbations here. 
It is noted that we also have another gauge choice 
where the metric perturbations become $C^0$ 
at the particle's location~\cite{Lousto:2008vw}.

%%%%%%%%%%%%%%%%%%%%%%%%%%%%%%
\subsection{Second order $\ell=2,\,m=\pm1$, odd parity perturbation}\label{subsec:SO21O}
%%%%%%%%%%%%%%%%%%%%%%%%%%%%%%

In the previous subsections, we focused on the first perturbative order. 
Here we treat the second perturbative order using the SRWZ formalism. 
We focus on the coupling between 
the first order $\ell=1,\,m=0$ even parity mode (with parity $(-1)^{1}$)
and the spin effect of the central black hole in Eq.~(\ref{eq:h0S}),
which is given by the $\ell=1,\,m=\pm1$ odd parity modes
 (with parity $(-1)^{1+1}$). 
This coupling creates the second order $\ell=2,\,m=\pm1$ odd parity
 perturbation (with parity $(-1)^{2+1}$). 

The Regge-Wheeler function for this odd parity perturbation
 satisfies 
\begin{eqnarray}
\left[
-{\frac {\partial ^{2}}{\partial {t}^{2}}}
+\frac{\partial^2}{{\partial r^*}^2}
-6\,{\frac {( r-2\,M) 
( {r} - {M}) }{{r}^{4} }}
\right] \,
\psi^{(2)\rm{(odd)}}_{2\pm1}(t,r)
= {\cal S}^{(2)\rm{RW}}_{2\pm1}(t,r) \,.
\end{eqnarray}
The source term is derived from the effective stress-energy tensor, 
\begin{eqnarray}
T_{\mu\nu}^{\rm (2,eff)} &=& 
- \frac{1}{8 \, \pi}G_{\mu\nu}^{(2)}[h^{\rm (1,dipole)}, h^{\rm (1,spin)}] \,,
\end{eqnarray}
by using the same tensor harmonics expansion 
as for the first perturbative order. 
When we consider the leading order in the slow motion approximation, 
the source term ${\cal S}^{\rm{RW}}_{2\pm1}$ becomes
\begin{eqnarray}
\fl
{\cal S}^{(2)\rm{RW}}_{2\pm1} = \pm \frac{\sqrt{10}\pi \mu S_x}{5M}
Y_{10}^{*}(0,0)
\nonumber \\ \times
\left(
\frac{2}{R(t)^2} \delta \left( r-R(t)  \right) 
- \frac{4}{R(t)} \frac{d}{dr}\delta \left( r-R(t)  \right) 
+ \frac{d^2}{dr^2}\delta \left( r-R(t)  \right)
\right) \,,
\end{eqnarray}
if we use the first order $\ell=1,\,m=0$ even parity mode 
under the D gauge. 
From this source term, we obtain the Regge-Wheeler function 
\begin{eqnarray}
\psi^{(2){\rm (odd)}}_{2\pm1}(t,r) &=& \pm \frac{4\sqrt{10}\pi}{15} 
\mu \frac{S_2}{R^2} \,Y_{10}^{*}(0,0)
\,,
%&=& \mp \frac{4\sqrt{10}\pi}{15} 
%\frac{\mu S_2}{M} \ddot R \,Y_{10}^{*}(0,0)
%\,,
\label{eq:psi2o21}
\end{eqnarray}
by using the same Green's function method discussed in 
Subsection~\ref{subsec:FO2030E}. Here, we have written $S_2=S_x$.

%%%%%%%%%%%%%%%%%%%%%%%%%%%%%%
\subsection{Spinning particle orbiting around a black hole}
%%%%%%%%%%%%%%%%%%%%%%%%%%%%%%

In the above subsection, 
we considered the black hole with mass $M$ that has 
a spin along the $x$-direction.
Here, we introduce a particle's spin which is parallel to 
the black-hole spin. 

First, for simplicity, 
we consider a point particle with mass $\mu$ 
that has a spin vector ${\bf S^{(\mu)}} = \{S_1,\,0,\,0\}$, 
and is located at ${\bf x_0} = \{0,\,0,\,R\}$ 
in the Cartesian coordinates. 
When we discuss perturbations from the spinning particle, 
we use the following energy-momentum tensor. 
\begin{eqnarray}
\fl 
T^{\alpha\beta} = T^{\alpha\beta}_{\rm mass} + T^{\alpha\beta}_{\rm spin} \,;
\nonumber \\ 
\fl 
T^{\alpha\beta}_{\rm mass} = \mu \int d\tau u^{\alpha}u^{\beta} 
\frac{\delta^{(4)}(x-z(\tau))}{\sqrt{-g}} \,,
\quad  
T^{\alpha\beta}_{\rm spin} = 
- \nabla_{\gamma} \int d\tau S_{(\mu)}^{\gamma (\alpha}u^{\beta)} 
\frac{\delta^{(4)}(x-z(\tau))}{\sqrt{-g}} 
\,,
\label{eq:Tmunu}
\end{eqnarray}
where we impose a spin supplementary 
condition (SSC) which determines the center of mass of the particle, 
$S_{(\mu)}^{\alpha\beta} u_{\beta}=0$. 
Since we focus on the leading order effect of the particle's spin, 
we may consider only the contribution of $T^{\alpha\beta}_{\rm spin}$. 
Furthermore, we can reduce this energy momentum tensor to the form
\begin{eqnarray}
T^{jt}_{\rm spin} = T^{tj}_{\rm spin}  
=  - \frac{1}{2} \int d\tau  \frac{1}{\sqrt{-g}} 
\partial_i [ S_{(\mu)}^{ij} \delta^{(4)}(x-z(\tau )) ] \,,
\end{eqnarray}
in the leading order of the slow motion approximation. 
The other components are higher order. 
Here, the spin tensor $S_{(\mu)}^{ij}$ is given by 
\begin{eqnarray}
S^{r\theta} = - \frac{\sin \phi}{r} S_1 \,, 
\quad 
S^{r\phi} =  - \frac{\cos \theta \cos \phi}{r \sin \theta} S_1 \,, 
\quad 
S^{\theta \phi} =  \frac{\cos \phi}{r} S_1 \,. 
\end{eqnarray}
The tensor harmonics coefficients of the stress-energy tensor are 
calculated as 
\begin{eqnarray}
\fl
A_{1\,\ell m}^{\rm spin}= \frac{i}{\sqrt{2}} \frac{S_1}{r^3}
\left(\sin \Phi \partial_{\theta} Y_{\ell m}^* (\Theta,\Phi)
+ \frac{\cos \Theta \cos \Phi}{\sin \Theta} \partial_{\phi} Y_{\ell m}^*(\Theta,\Phi) \right) 
\delta (r-R) \,,
\nonumber \\ 
\fl 
B_{0\,\ell m}^{\rm spin} = \frac{i}{\sqrt{2\ell(\ell+1)}} \frac{S_1}{r}
\left(\sin \Phi \partial_{\theta} Y_{\ell m}^* (\Theta,\Phi)
+ \frac{\cos \Theta \cos \Phi}{\sin \Theta} \partial_{\phi} Y_{\ell m}^*(\Theta,\Phi) \right) 
\partial_r \left(\frac{\delta (r-R)}{r}\right) \,,
\nonumber \\ 
\fl 
Q_{0\,\ell m}^{\rm spin} = -\frac{1}{\sqrt{2\ell(\ell+1)}} \frac{S_1}{r}
\left(\frac{\sin \Phi}{\sin \Theta} \partial_{\phi} Y_{\ell m}^* (\Theta,\Phi)
- \cos \Theta \cos \Phi \partial_{\theta} Y_{\ell m}^*(\Theta,\Phi) \right) 
\nonumber \\ \times 
\partial_r \left(\frac{\delta (r-R)}{r}\right) \,,
\end{eqnarray}
where we set the location of the particle on the $z$-axis. 
It should be noted that 
only $\ell \geq 1$, $m=\pm1$ modes have a non-zero value 
because of $Y_{\ell m}^*(\Theta,\Phi) \sim (\sin \Theta)^{|m|}$. 
And also, both the even and odd parity modes 
exist in the metric perturbations from the particle's spin. 

We evaluate the $\ell=2,\,m=\pm1$ odd parity 
perturbations from $T^{\alpha\beta}_{\rm spin}$ 
in the first perturbative order calculation. 
These perturbations have the leading order effect of the particle's spin. 
The tensor harmonics coefficient of the stress-energy tensor is 
given by 
\begin{eqnarray}
Q_{0\,2 \pm1}^{\rm spin} &=& \mp \sqrt{\frac{5}{32\pi}} 
\frac{S_1}{r} \partial_r \left(\frac{\delta (r-R)}{r}\right) \,.
\end{eqnarray}
The wave function is obtained from the Regge-Wheeler equation, 
\begin{eqnarray}
\left[
-{\frac {\partial ^{2}}{\partial {t}^{2}}}
+\frac{\partial^2}{{\partial r^*}^2}
-6\,{\frac {( r-2\,M) 
( {r} - {M}) }{{r}^{4} }}
\right] \,
\psi^{(1){\rm (odd)}}_{2\pm1}(t,r)
= {\cal S}^{\rm RW,spin}_{2\pm1}(t,r) \,, 
\end{eqnarray}
and the source term in the leading order 
of the slow motion approximation is given by 
\begin{eqnarray}
{\cal S}^{\rm RW,spin}_{2\pm1} = \mp \sqrt{\frac{5\pi}{6}} S_1
\left(
- \frac{2}{R(t)} \frac{d}{dr}\delta \left( r-R(t)  \right) 
+ \frac{d^2}{dr^2}\delta \left( r-R(t)  \right)
\right) \,,
\end{eqnarray}
where we have extended the interpretation of the particle's location to
a sequence of quasistatic locations $R=R(t)$. From the above source term, 
we obtain the Regge-Wheeler function
\begin{eqnarray}
\psi^{(1){\rm (odd)}}_{2\pm1}(t,r) = \pm \frac{\sqrt{120\pi}}{15} 
S_1 \ddot R 
\,,
\label{eq:psi1o21}
\end{eqnarray}
by using the same Green's function discussed in Subsection~\ref{subsec:FO2030E}. 
This wave function has the following relation with 
the black-hole spin effect in Eq.~(\ref{eq:psi2o21}). 
\begin{eqnarray}
\psi^{(1){\rm (odd)}}_{2\pm1}(t,r) = 
- \frac{S_1}{S_2}\frac{M}{\mu} \psi^{(2){\rm (odd)}}_{2\pm1}(t,r) 
\,.
\label{eq:psi21_sym}
\end{eqnarray}

%%%%%%%%%%%%%%%%%%%%%%%%%%%%%%
\subsection{Evolution of the linear momentum}\label{subsec:ELMs}
%%%%%%%%%%%%%%%%%%%%%%%%%%%%%%

Using the results in the previous subsections 
and the approximated equation of motion 
in the slow motion approximation, $d^2R/dt^2 = -M/R^2$, 
we can calculate the leading order evolution 
of the linear momentum. 
The contributions of the gravitational waveforms are summarized 
in Table~\ref{table:Contrib}. The mass, velocity, orbital radius and spin dependence 
is estimated by 
\begin{eqnarray*}
\frac{v^2}{R^2} \,\psi^{(n)}_{\ell m} \, \psi^{(n')}_{\ell'm'} 
\,,
\end{eqnarray*}
from Eq.~(\ref{eq:dotPi}). Also, the direction can be 
derived from the angular integrations in Eq.~(\ref{eq:PsPc}). 

\begin{table}
\caption{Leading order mode contributions to the evolution 
of the linear momentum
in the head-on collision of spinning black holes.}
\label{table:Contrib}
\begin{center}
\begin{tabular}{l|l|c}
\hline
combination & dependence & direction \\
\hline
($\ell=2,\,m=0$, even) $\cdot$ ($\ell=3,\,m=0$, even)    & $\mu^2 v^7/R^2$     & z     \\
($\ell=2,\,m=0$, even) $\cdot$ ($\ell=2,\,m=\pm 1$, odd) & $\mu S_1 v^6/R^3$   & x-y   \\
($\ell=2,\,m=0$, even) $\cdot$ ($\ell=2,\,m=\pm 1$, odd) & $\mu^2 S_2 v^4/R^4$ & x-y   \\
\hline
\end{tabular}
\end{center}
\end{table}

First, we find 
\begin{eqnarray}
{\dot P}_x = 0 \,.
\end{eqnarray}
This result is also obtained by analyzing the symmetry of the system. 
In practice, we may use the symmetry between $m$ and $-m$ modes 
in the coefficients of the tensor harmonics of the metric perturbation  
$G^{\rm AF(i)}_{\ell m}$ and $h^{\rm AF(i)}_{2\,\ell m}$, 
and the integration $P^C_i$ and $P^S_i$. 

Next, we discuss the spin independent contribution to the evolution 
of the linear momentum. This contribution arises in the $z$-direction as 
\begin{eqnarray}
{\dot P}_z = - \frac{16}{105} \frac{\mu^2 M^2}{R^4}
\left({\dot R}^3 - \frac{2M}{R} {\dot R} \right)
 \,.
\label{eq:Pz2}
\end{eqnarray}
The above equation is derived from the combination 
of the first order $\ell=2$, $m=0$ and $\ell=3$, $m=0$ even parity modes 
in Eqs.~(\ref{eq:psi20e}) and (\ref{eq:psi30e}). 

Finally, for the $y$-direction, we have 
the leading order contribution of the spin effects 
for the evolution of the linear momentum.  
\begin{eqnarray}
{\dot P}_y = - \frac{16}{15} \,\mu^2 \,M^2 
\frac{{\dot R}^2}{R^5} \left(\frac{S_2}{M} - \frac{S_1}{\mu}\right) \,.
\label{eq:Py12}
\end{eqnarray}
This is calculated from the combination of $\ell=2$, $m=0$ even parity 
and the $\ell=2$, $m=\pm1$ odd parity modes 
which include two different wave functions given in 
Eqs.~(\ref{eq:psi2o21}) and~(\ref{eq:psi1o21}). 

These results are consistent with the calculation 
in the post-Newtonian approach~\cite{Kidder:1995zr}. 
Note that Kidder~\cite{Kidder:1995zr} derived 
the above results for general orbits. 
On the other hand, our calculation discussed here 
is limited to the head-on collision of spinning black holes 
given in Subsection~\ref{subsec:PPM}.

%%%%%%%%%%%%%%%%%%%%%%%%%%%%%%%%%%%%%%%%%%%%%%%%%%%%%%%%%%%%%%%%%%%%%%%%
\section{Discussion} 
%%%%%%%%%%%%%%%%%%%%%%%%%%%%%%%%%%%%%%%%%%%%%%%%%%%%%%%%%%%%%%%%%%%%%%%%

We have discussed gravitational wave recoil effects 
and analytically derived the leading order effects in the evolution 
of the linear momentum by using the SRWZ formalism. 
This formalism is an extension of the RWZ formalism 
with a perturbative spin of the background black hole. 

In the appendix of~\cite{Lousto:2010qx}, 
we applied the perturbative spin formalism
to compute the corresponding quasi-normal modes 
and compare them with those obtained for the Kerr black hole 
for all values of the spin parameter. 
These results show that the SRWZ formalism provide reliable predictions 
for the spin parameter $a/M\leq0.3$. 

From the analytic treatment of the gravitational radiation recoil,  
we confirm the leading $q^2$ (where $q=\mu/M$) dependence of the
large recoils out of the orbital plane~\cite{Lousto:2008dn}. 
In the black hole perturbation theory, this scaling 
with mass ratio is trivial, and in~\cite{Sundararajan:2010sr} 
the spin dependence has also been discussed in detail 
by solving Teukolsky equation~\cite{Teukolsky:1973ha} numerically. 

The results of the recoil obtained here can be extended 
to higher PN order in the sense of the slow motion approximation 
where the characteristic orbital velocity 
$v \ll 1$~(for example, see~\cite{Sasaki:2003xr}). 
On the other hand, higher order spin effects are complicated. 
In this paper, we have focused only on the couplings between 
the first order perturbations about a Schwarzschild background 
and the black-hole spin. However, it is also necessary 
to treat the equations of motion with spin. 
We may use the Teukolsky formalism~\cite{Teukolsky:1973ha} 
for the spinning large black hole, i.e., use a Kerr background. 
Taking into account the spin of the particle adds a new degree of
 complication that makes it difficult to obtain an analytic expression 
of the gravitational radiation recoil for general orbits 
and we would need to perform a numerical calculation.

%%%%%%%%%%%%%%%%%%%%%%%%%%%%%%%%%%%%%%%%%%%%%%%%%%%%%%%%%%%%%%%%%%%%%%%%
\section*{Acknowledgments}
%%%%%%%%%%%%%%%%%%%%%%%%%%%%%%%%%%%%%%%%%%%%%%%%%%%%%%%%%%%%%%%%%%%%%%%%

We gratefully acknowledge the NSF for financial support from Grants
No. PHY-0722315, No. PHY-0653303, No. PHY-0714388, No. PHY-0722703,
No. DMS-0820923, No. PHY-0929114, No. PHY-0969855, No. PHY-0903782,
No. CDI-1028087; and NASA for financial support from NASA Grants
No. 07-ATFP07-0158 and No. HST-AR-11763.

%%%%%%%%% references %%%%%%%%%%%%%%%%%%%%%%%%%%%%%%

\end{document}